\documentstyle[12pt]{article}
\def\lapproxeq{\lower .7ex\hbox{$\;\stackrel{\textstyle <}{\sim}\;$}}
\def\qbar{\bar q}
\def\ubar{\bar u}
\def\dbar{\bar d}

\begin{document}
\begin{titlepage}
\vspace*{-1cm}
\begin{flushright}
DTP/93/12 \\
RAL-93-014\\
March 1993
\end{flushright}
\vskip 1.cm
\begin{center}
{\Large\bf Phenomenological constraints on the flavour\\
asymmetry of the nucleon sea
}
\vskip 1.cm
{\large A.D. Martin and W.J. Stirling}
\vskip .2cm
{\it Department of Physics, University of Durham \\
Durham DH1 3LE, England }\\
\vskip .4cm
and
\vskip   .4cm
{\large R.G. Roberts}
\vskip .2cm
{\it
Rutherford Appleton Laboratory,  \\
Chilton, Didcot  OX11 0QX, England
} \\
\vskip 1cm
\end{center}

\begin{abstract}
We study the possible flavour asymmetry, $\bar{u} \neq \bar{d}$, of the light
quark sea distributions of the proton.  We discuss the information that is at
present available from data on deep-inelastic lepton-nucleon scattering
and from Drell-Yan production on various nuclear targets.
We show that the ratio
of dilepton yields on hydrogen and deuterium targets is very sensitive to
$\bar{u}-\bar{d}$.
\end{abstract}

\vfill
\end{titlepage}
\newpage

The increased precision of deep inelastic lepton-nucleon scattering
data has considerably improved our knowledge of the parton distributions of
the proton.  Until recently all global structure function analyses assumed
that the light quark sea distributions satisfied the equality $\bar{u} =
\bar{d}$.  However, motivated by the improvement of the data, particularly at
small $x$, and by the consequences for the Gottfried sum rule, analyses
\cite{MRS1,MRS2,CTEQ} have been performed with $\bar{u} \neq \bar{d}$.  Here
we wish to explore how large an inequality the available data can tolerate and
to see how well future experiments can pin down $\bar{u}-\bar{d}$.

Valuable insight into the problem is obtained by expressing (to leading order)
the observed muon and neutrino deep inelastic structure functions in terms of
the parton densities
\begin{eqnarray}
F^{\mu p}_2 - F^{\mu n}_2 \;\; &=& \;\; {\textstyle
 \frac{1}{3}} x(u+\bar{u}-d-\bar{d})
\\
{\textstyle\frac{1}{2}}
(F^{\mu p}_2 + F^{\mu n}_2) \;\; &=& \;\; {\textstyle\frac{5}{18}}
x(u+\bar{u}+d+\bar{d}+{\textstyle\frac{4}{5}}s)
\\
F^{\nu N}_2 \;\; &=& \;\; F^{\bar{\nu}N}_2 \; = \; x(u+\bar{u}+d+\bar{d}+2s)
\\
{\textstyle\frac{1}{2}}x(F^{\nu N}_3 + F^{\bar{\nu}N}_3) \;\; &=& \;\;
x(u-\bar{u}+d-\bar{d})  \; ,
\end{eqnarray}
 where $N$ is an isoscalar nuclear target, and where we have
included the $s$ quark, but neglected the (small) $c$ quark, contribution.
These four observables determine four combinations of parton densities, which
we can take to be $u+\bar{u}, \, d+\bar{d}, \, \bar{u}+\bar{d}$ and $s$.  In
other words data at a given $x,Q^2$ cannot verify whether $\bar{u} = \bar{d}$
or not.

The above discussion is very simplistic because in practice we must (i)
include the next-to-leading $O(\alpha_s)$ contributions, which require
knowledge of the gluon distribution, (ii) allow for possible adjustments in
the overall relative normalisation of the data sets, (iii) include the heavy
target corrections of the neutrino data, and (iv) include screening
corrections to allow $F^{\mu n}_2$ to be obtained from deuterium data.
Assuming that this has all been done correctly, we compare in Fig.\ 1 the
parton distributions at $Q^2 = 20$ GeV$^2$ from the two most recent global
analyses \cite{MRS2,CTEQ}.  If we make a detailed comparison at, say, $x =
0.15$ then we see significant differences between $u,d,\bar{u},\bar{d}$ from
the two sets of partons.  For CTEQ \cite{CTEQ} we have $\bar{d} \simeq
2\bar{u}$, while for MRS \cite{MRS2} we have $\bar{d} \simeq
\bar{u}$.\footnote{This difference is a consequence of the flexible (and
independent) parametrizations of the $\bar{u},\bar{d},s$ distributions allowed
in the CTEQ analysis \cite{CTEQ}.  However the most dramatic difference
between the CTEQ and MRS distributions is in the behaviour of the $s$ quark
distribution (the dash-dot curves on Fig.\ 1).  The freely parametrized CTEQ
$s$ distribution becomes relatively large at small $x$ to accommodate the
$F^{\nu N}_2$ data for $x \lapproxeq 0.1$.  On the other hand MRS
\cite{MRS1,MRS2} adopted a less flexible parametrization, with $s =
\frac{1}{4}(\bar{u}+\bar{d})$ at $Q^2 = 4$ GeV$^2$, and allowed for
uncertainties in the heavy target corrections to the neutrino data; the
(factor of 2) suppression of the $s$ quark distribution was chosen so as to be
in better accord with dimuon production in deep inelastic neutrino scattering
\cite{CCFR}.}  However this difference is compensated by $u$(CTEQ) $> u$(MRS)
and $d$(CTEQ) $< d$(MRS) so that the observable combinations $u+\bar{u}, \,
d+\bar{d}$ and $\bar{u}+\bar{d}$ remain essentially the same in both analyses,
as indeed they should be for values of $x$ where accurate data exist.  So, as
mentioned above, without additional input, the structure function data on
their own do not test $\bar{u} \neq \bar{d}$.

The motivation for $\bar{u} \neq \bar{d}$ comes from the NMC measurement
\cite{NMC}
\begin{equation}
\int^{0.8}_{0.004} \frac{dx}{x} (F^{\mu p}_2 - F^{\mu n}_2) \;\; = \;\; 0.227
\pm 0.007 ({\rm stat.}) \pm 0.014 ({\rm sys.})
\end{equation}
at $Q^2 = 4$ GeV$^2$, as compared to the Gottfried sum rule
\begin{eqnarray}
I_{GSR} \; = \; \int^1_0 \frac{dx}{x} (F^{\mu p}_2 - F^{\mu n}_2) \; &=& \;
\frac{1}{3} \int^1_0 dx(u_v-d_v) + \frac{2}{3} \int^1_0 dx(\bar{u}-\bar{d})
\nonumber \\
&=& \; \frac{1}{3} \hspace*{1cm} {\rm if} \hspace*{.75cm} \bar{u} = \bar{d}\; ,
\end{eqnarray}
  where here $u$ and $d$ have been expressed as the sum of valence
and sea distributions, {\it e.g.}\ $u = u_v + u_s$ and $\bar{u} = u_s$.  A
straightforward comparison of (5) and (6) implies that $\bar{d} > \bar{u}$,
and indeed from the lack of Regge $f_2-a_2$ exchange degeneracy we expect
\begin{equation}
\bar{d} - \bar{u}  \;\; = \;\; Ax^{-\alpha_R} (1-x)^{\eta_S}
\end{equation}
with the small $x$ behaviour governed by the Regge (meson
trajectory) intercept $\alpha_R \simeq 0.5$.  In the global analyses of
refs.~\cite{MRS1,MRS2} we parametrized $\bar{d}-\bar{u}$ in this way and found
$I_{GSR} \simeq 0.24-0.26$.  However it is possible to achieve an equally
acceptable global fit of all the structure function data with $\bar{u} =
\bar{d}$ (and hence $I_{GSR} = \frac{1}{3})$, albeit with a contrived small
$x$ behaviour of the valence distributions \cite{MRS1}.

Fig.\ 2 shows the ingredients of the Gottfried sum rule for various sets of
parton distributions with $\bar{u} \neq \bar{d}$.  The central part of the
figure, which shows $F^{\mu p}_2-F^{\mu n}_2$ on a logarithmic $x$ scale, is a
visual display of the integrand of the sum rule at $Q^2 = 7$ GeV$^2$.  The
data are from NMC \cite{KAB}.  The top part of the figure shows the
accumulated contribution to the sum rule as we integrate down from $x$ = 1;
thus the limiting values of the curves as $x \rightarrow 0$ give $I_{GSR}$ of
(6).  The central curves in the shaded bands are obtained from the
MRS(D$^{\prime}_0$) partons with $\bar{d}-\bar{u}$ parametrized
at $Q^2 = 4$ GeV$^2$ as in (7); these
partons give $I_{GSR}$ = 0.256 \cite{MRS2}.

In order to explore the allowed variations in the {\it shape} of the $x$
behaviour of $\bar{d}-\bar{u}$ we have repeated the global analysis of
ref.~\cite{MRS2} with the difference having the extended parametrization
\begin{equation}
\bar{d}-\bar{u} \;\; = \;\; Ax^{-\alpha_R}(1-x)^{\eta_S} (1+\gamma x + \delta
x^2) ,
\end{equation}
 with $A$ chosen to maintain the normalisation $I_{GSR}$ = 0.256.
First we set $\delta = 0$.  We find that we can maintain the quality of the
fit to the data, that was achieved by the D$^{\prime}_0$ fit of
ref.~\cite{MRS2}, provided that
$\gamma$ lies in the range $-8 \lapproxeq \gamma \lapproxeq 32$.  These
extreme \lq\lq hard" and \lq\lq soft" versions of $\bar{d}-\bar{u}$ of
D$^{\prime}_0$ correspond to the boundary curves of the shaded regions shown
in Fig.\ 2.  We may thus regard the shaded region as a first qualitative
estimate of the uncertainty in $\bar{d}-\bar{u}$.  We see that this range of
fits only modifies $F^{\mu p}_2-F^{\mu n}_2$ for $x \lapproxeq 0.05$;
indeed deep-inelastic data impose only a weak constraint on $\bar{d}-\bar{u}$
at larger values of $x$.  For a determination of $\bar{d}-\bar{u}$ at larger
$x$ we should compare Drell-Yan production of $pp$ and $pn$ origin.  A hint of
what may be possible is shown in Fig.\ 3, which compares Drell-Yan data
\cite{E772} obtained from a neutron rich (tungsten) target with those from
isoscalar (deuterium and carbon)  targets.  We discuss such Drell-Yan
comparisons, in detail, below.

We have also repeated the global analysis of the deep inelastic data with
$\delta
\neq 0$ in the parametrization (8) of $\bar{d}-\bar{u}$, in an attempt to
reproduce the qualitative trend of the Drell-Yan data of Fig.\ 3.  The
predictions of a fit with $\gamma = -60, \, \delta = 300$ are shown by the
dashed curves on Figs.\ 2 and 3.  These have a radically different shape and,
in fact, the quality of the fit to the deep inelastic data diminishes; we
regard this as an example of extreme behaviour of $\bar{d}-\bar{u}$.  For
completeness we show by dot-dashed curves the predictions of the CTEQ1M
analysis, which is in the spirit of the D$^{\prime}_0$ fit but with
$\bar{u},\bar{d},s$ much more freely parametrized.  We see a dramatically
different behaviour at small $x$; indeed $F^{\mu p}_2-F^{\mu n}_2$ becomes
negative for $x < 0.006$ and as a result the value of $I_{GSR}$ is well below
0.2.

The distributions shown in the lower part of Fig.\ 2 therefore offer (i) a set
of conservative variations of $\bar{d}-\bar{u}$ of the D$^{\prime}_0$ fit
which are consistent with all the deep inelastic data, together with (ii) two
examples of unusual (but interesting) behaviour which arise if the sea
distributions are more freely parametrized.

The Drell-Yan process provides a complementary method of investigating the
antiquark content of the nucleon. In leading order, for a proton beam on
a nuclear target $A$,
\begin{equation}
{d^2\sigma \over d M^2 d x_F} = {4\pi \alpha^2 \over 9 M^2 s}\;
 {1\over x_p + x_A  }\;
\sum_q e_q^2 \; [q(x_p) \qbar(x_A) + \qbar(x_p) q(x_A) ] \; ,
\end{equation}
with $\tau = M^2/s =  x_p x_A$ and $x_F = x_p - x_A$.
The crucial point is that a comparison of the cross sections on targets
with {\it different} numbers of protons and neutrons leads directly to
information on $\dbar - \ubar$. Two methods have been suggested.
The first \cite{E772} relies on the fact  that at large $x_F$
 the cross section (9)  is dominated by the annihilation of $u$ quarks
in the proton with $\ubar$ quarks in the target. If the latter contains
$Z$ protons and $A-Z$ neutrons, then
\begin{equation}
{d^2\sigma \over d M^2 d x_F}\;  \approx\;  {4\pi \alpha^2 \over 9 M^2 s} \;
{1\over x_p + x_A}\; {4\over 9} \;
 u(x_p) {1\over A} \{ Z  \ubar(x_A) + (A-Z)  \dbar(x_A) \} \;
{}.
\end{equation}
In particular the ratio of non-isoscalar to isoscalar target cross sections
 is \cite{E772}
\begin{equation}
R_{A/IS}(x)\;  \approx\;  1\;  +\;
\left( 1 - {2Z\over A} \right)\; {\dbar(x)  - \ubar(x) \over
\dbar(x) + \ubar(x) }  \; ,
\label{rwis}
\end{equation}
with  $ x \equiv x_A$.
The ratio for tungsten (with $1-2Z/A = 0.195$) to isoscalar  targets
 has recently been measured by the E772
collaboration \cite{E772}. A comparison of  their data with the predictions of
the
various parton sets discussed above is  shown in Fig.~3. The open circles
at small $x$ show the data before a correction for nuclear shadowing is
applied \cite{E772}. It turns out that the $x_F $ values are largest
for the data points at small $x$, and it is only for these that the
approximation (\ref{rwis}) is valid. At larger $x$, the smaller $x_F$
values lead to a contamination of the $u(x_p) \bar u(x_A)$ annihilation with
the other processes, and a corresponding decrease of the sensitivity
to $\dbar - \ubar$.
The shaded band in Fig.~3 again corresponds to the range $\delta = 0$,
$-8 < \gamma < 32$ of the $\dbar - \ubar$ parametrization of (8),
evaluated using the full Drell-Yan next-to-leading order cross section.
The D$'_0$ prediction (solid line) is roughly in the middle of this band.
The S$'_0$ prediction (not shown)
has $R \simeq  1$ for small $x$, falling slightly
to $R\simeq 0.99$ at the larger $x$ values.  All these predictions are
clearly consistent with the measured ratio.
Notice that there is perhaps some evidence  that $R < 1$ ({\it i.e.}
$\dbar < \ubar$)  for $x \lapproxeq 0.15$.  As mentioned above,
it is not difficult to derive a
$\dbar - \ubar  $ form which follows this trend, as shown by the dashed line
in Fig.~3 which corresponds to $\delta = 300$, $\gamma = -60$,
though this parametrization leads to a poorer overall fit to the deep inelastic
data.
Note also that the CTEQ1M prediction (dot-dashed curve), which has
$\dbar - \ubar$ large and positive for the range of  $x$ spanned by the data,
 appears to be  slightly disfavoured.

A much more powerful Drell-Yan measurement is to compare the cross sections
on proton and neutron (in practice hydrogen and deuterium)
 targets \cite{SDEWJS}.
Two such experiments have recently been proposed \cite{CERNDY,FNALDY}.
Fig.~4 shows the next-to-leading order QCD prediction for the asymmetry
\begin{equation}
A_{DY} = { \sigma(pp) - \sigma(pn)
   \over \sigma(pp) + \sigma(pn) }   \; ,
\end{equation}
where $\sigma \equiv  d^2 \sigma/dM dy|_{y=0}$,
as a function of $\sqrt{\tau}$ at $p_{\rm beam} = 450 $ GeV/c,
corresponding to the proposed CERN experiment \cite{CERNDY}.\footnote{The
predictions for the corresponding Fermilab energy  are
virtually indistinguishable.} The parton
distributions are the same as in Fig.~3. Because $u > d$ in the proton,
 the asymmetry
is positive for sets with $\dbar -\ubar $ zero or small at large $x$.
The asymmetry is reduced and can even become negative for sets with
large $\dbar - \ubar$, in particular the CTEQ1M set. A measurement of
this asymmetry in the $\sqrt{\tau} = 0.1 - 0.3 $ range to an accuracy
of a few percent will therefore provide a powerful discriminator
of the possible behaviours of the $\dbar - \ubar$ difference.
Note also that the sensitivity of the asymmetry (12)
to $\dbar - \ubar $ increases at forward
rapidity (equivalently $x_F > 0$): at large $y$,  where $u \ubar $ annihilation
again dominates, we have simply that $A_{DY} \approx (\ubar - \dbar) / (\ubar
+ \dbar) $, which is equivalent to (\ref{rwis}) with $1-2Z/A = 1$.

In conclusion, we find that deep inelastic data do not unambiguously
determine $\ubar$ and $\dbar$. However, if a reasonable behaviour is imposed
on $\dbar - \ubar$ at small $x$ (motivated by a small breaking of meson
exchange degeneracy) then the data indicate a small flavour asymmetry with
$\dbar > \ubar $ for $x \lapproxeq 0.01$. Global fits to data imply that the
$\dbar > \ubar$ inequality persists to larger $x$, but it is possible to
introduce
more flexible parametrizations of $\dbar - \ubar$. We show that
a comparison of Drell-Yan production on hydrogen and deuterium targets
will provide a good  determination of $\dbar - \ubar$ in the medium $x$ range.
 Finally,
it is worth noting that
a small, positive $\dbar - \ubar$ is a common feature of most model
calculations
\cite{OTHERS,EHQ} of the sea quark flavour
asymmetry, in particular those involving mesonic contributions to the
deep inelastic structure functions. For example, the Drell-Yan
predictions of the chiral quark model
of ref.~\cite{EHQ} are very similar to those of our standard D$'_0$ fit.

\newpage
\noindent{\Large\bf Figure Captions}

\begin{itemize}
\item[{[1]}] The parton distributions of the proton at $Q^2 = 20$ GeV$^2$
corresponding to (a) set D$^{\prime}_0$ of MRS \cite{MRS2} and to (b) set
CTEQ1M of \cite{CTEQ} .

\item[{[2]}] The upper plot shows the accumulated contribution to the
Gottfried sum rule, (6), as a function of $x$, the lower limit of integration,
which is obtained from sets of partons with differing behaviour of
$\bar{d}-\bar{u}$, as shown in the lower plot (for $Q^2 = 7$ GeV$^2$).  The
shaded bands contain a family of parton solutions with $\bar{d}-\bar{u}$ given
by (8) with $-8 < \gamma < 32$ and $\delta = 0$; the central curves in the
bands corresponds to set D$^{\prime}_0$ of \cite{MRS2} with $\gamma = 0$.  The
dashed curves correspond to a fit with $\delta \neq 0$ in (8), and the
dot-dashed curves correspond to set CTEQ1M of \cite{CTEQ}.  The central part
of the figure shows the integrand of the Gottfried sum rule, together with NMC
data \cite{KAB} which we have corrected, at small $x$, for deuteron screening
effects.

\item[{[3]}] The ratio of dilepton yields per nucleon from tungsten and
isoscalar (deuterium and carbon)  targets as a function of $x$(target).
 The data are from ref.~\cite{E772};
the open circles at small $x$ show the ratio before the
correction for nuclear shadowing. The curves are the full next-to-leading
order QCD predictions using the same parton sets as in Fig.~2. Thus
the shaded band corresponds to a range of distributions
 with $\bar{d}-\bar{u}$ given
by (8) with $-8 < \gamma < 32$ and $\delta = 0$; the central curve in the
band corresponds to set D$^{\prime}_0$ of \cite{MRS2} with $\gamma = 0$.  The
dashed curve corresponds to a fit with $\delta = 300,\
 \gamma= -60$ in (8), while the
dot-dashed curve corresponds to set CTEQ1M of \cite{CTEQ}.

\item[{[4]}]
Predictions  for the Drell-Yan asymmetry
$ A_{DY} = ( \sigma(pp) - \sigma(pn) )
/( \sigma(pp) + \sigma(pn) ) $,
where $\sigma \equiv  d^2 \sigma/dM dy|_{y=0}$, as a function
of $\sqrt{\tau}= M/\sqrt{s}$ with
$p_{\rm beam} = 450 $ GeV/c.  The curves are based on next-to-leading
order calculations using the same parton distributions as in Fig.~3.

\end{itemize}

\end{document}